\documentstyle[twocolumn,prl,aps]{revtex}
\begin{document}
\draft
\title{An arbitrary-operation gate with a SQUID qubit}
\author{Chui-Ping Yang \thanks{%
Email address: cpyang@ku.edu} and Siyuan Han \thanks{%
Email address: han@ku.edu}}
\address{Department of Physics and Astronomy, University of Kansas, Lawrence, Kansas\\
66045}
\maketitle

\begin{abstract}
A novel approach is proposed for realizing an arbitrary-operation gate with
a SQUID qubit via pulsed-microwave manipulation. In this approach, the two
logical states of the qubit are represented by the two lowest levels of the
SQUID and an intermediate level is utilized for the gate manipulation. The
method does not involve population in the intermediate level or tunneling
between the two logical qubit states during the gate operation. Morever, we
show that the gate can be much faster than the conventional two-level gate.
In addition, to take the advantage of geometric quantum computing, we
further show how the method can be extended to implement an arbitrary
quantum logic operation in a SQUID qubit via geometric manipulation.
\end{abstract}



In the growing field of quantum computing, superconducting devices including
cooper pair boxes, Josephson junctions, and superconducting quantum
interference devices (SQUIDs) [1-10] have appeared to be among the most
promising candidates for quantum computing. Reasons for this are that these
systems can be easily fabricated to implement a large-scale quantum
computing and have been demonstrated to have relatively long decoherence
time [11,12]. In the past few years, for SQUID systems, many methods for
demonstrating macroscopic quantum coherence [7], creating a superposition
state (i.e., a Sh\"ordinger cat state) [9], or completing population
transfer between low-lying levels (i.e., a NOT gate) [10] have been
presented. However, the key point for quantum computing is how to realize an
arbitrary single-qubit operation plus a two-qubit controlled-NOT (or
controlled-phase shift) operation, since this set of operations makes a
universal quantum computing [13].

In this letter, we focus on how to achieve an arbitrary-operation gate with
a SQUID qubit. The proposed method has several advantages: (a) during the
gate operation, the intermediate level is not populated and thus decoherence
induced by spontaneous emission from the intermediate level is greatly
suppressed; (b) no tunneling between the SQUID-qubit levels $\left|
0\right\rangle $ and $\left| 1\right\rangle $ is required during the gate
operation, therefore the decay from the qubit levels can be made negligibly
small via increasing the potential barrier between the qubit levels; (c)
more importantly, as shown below, the gate operation can be performed faster
compared with the conventional gate operation. In addition, we show that the
approach can be extended to perform an arbitrary operation on a SQUID qubit
via geometric manipulation of phase.

Consider a SQUID driven by two classical microwave pulses $I$ and $II$. In
the following, the SQUID is treated quantum mechanically, while the
microwave pulses are treated classically. The Hamiltonian $H$ for the
coupled system can be written as 
\begin{equation}
H=H_s+H_i^{\left( I\right) }+H_i^{\left( II\right) },
\end{equation}
where $H_s$ is the Hamiltonian for the SQUID; and $H_i^{\left( I\right) }$
and $H_i^{\left( II\right) }$ are the interaction energies for the SQUID
with the two pulses, respectively.

The SQUIDs considered throughout this letter are rf SQUIDs each consisting
of a Josephson tunnel junction enclosed by a superconducting loop. The
Hamiltonian for an rf SQUID (with junction capacitance $C$ and loop
inductance $L$) can be written in the usual form 
\begin{equation}
H_s=\frac{Q^2}{2C}+\frac{\left( \Phi -\Phi _x\right) ^2}{2L}-E_J\cos \left(
2\pi \frac \Phi {\Phi _0}\right) ,
\end{equation}
where $\Phi $ is the magnetic flux threading the ring, $Q$ is the total
charge on the capacitor, $\Phi _x$ is the static (or quasistatic) external
flux applied to the ring, and $E_J$ $\equiv I_c\Phi _0/2\pi $ is the
Josephson coupling energy ($I_c$ is the critical current of the junction and 
$\Phi _0=h/2e$ is the flux quantum). In practice, the single junction is
often replaced by a dc SQUID with low inductance $l\ll L$ which is
effectively a junction with its critical current tunable by varying the flux
applied to the dc SQUID.

The SQUID is coupled inductively to the $k$th pulse ($k=I,II$) through the
interaction energy $H_i^k=\lambda _k\left( \Phi -\Phi _x\right) \Phi _{\mu
w}^k,$ where $\lambda _k=-1/L$ is the coupling coefficient between the SQUID
and the $k$th pulse; and $\Phi _{\mu w}^k$ is the magnetic flux generated by
the magnetic component ${\bf B}^k\left( t\right) $ of the $k$th pulse, which
is given by $\Phi _{\mu w}^k\equiv \int_S{\bf B}^k\left( t\right) \cdot d%
{\bf S}$ ($S$ is any surface that is bounded by the ring). The expression of 
${\bf B}^k(t)$ takes the form ${\bf B}^k\left( t\right) ={\bf B}_0^k{\bf %
\cos }\left( \omega _kt+\phi _k\right) ,$ where ${\bf B}_0^k,$ $\omega _k$
and $\phi _k$ are the amplitude, frequency and phase of the $k$th pulse,
respectively.

Consider a $\Lambda $-type configuration formed by the two lowest levels and
an intermediate level of the SQUID, denoted by $\left| 0\right\rangle
,\left| 1\right\rangle $ and $\left| a\right\rangle $ with energy
eigenvalues $E_0,E_1,$and $E_a,$ respectively [Fig. 1(a)]. Suppose that the
coupling of $\left| 0\right\rangle ,\left| 1\right\rangle $ and $\left|
a\right\rangle $ with the other levels via the pulses is negligible, which
can always be realized by adjusting pulse frequencies, or level spacings of
the SQUID. Under this consideration, it is easy to find that (i) if no $%
\left| 0\right\rangle \leftrightarrow \left| 1\right\rangle $ and $\left|
1\right\rangle \leftrightarrow \left| a\right\rangle $ transitions are
induced by the first pulse, i.e., the first pulse is far-off resonant with
the $\left| 0\right\rangle \leftrightarrow \left| 1\right\rangle $ and $%
\left| 1\right\rangle \leftrightarrow \left| a\right\rangle $ transitions,
which can be realized by setting $\left| \omega _{01}-\omega _I\right| \gg
\Omega _{01}^I$ and $\left| \omega _{1a}-\omega _I\right| \gg \Omega
_{1a}^I; $ (ii) if the second pulse is far-off resonant with $\left|
0\right\rangle \leftrightarrow \left| 1\right\rangle $ and $\left|
0\right\rangle \leftrightarrow \left| a\right\rangle $ transitions, i.e., $%
\left| \omega _{01}-\omega _{II}\right| \gg \Omega _{01}^{II}$ and $\left|
\omega _{0a}-\omega _{II}\right| \gg \Omega _{0a}^{II};$ and (iii) if the
detuning of the first pulse with the $\left| 0\right\rangle \leftrightarrow
\left| a\right\rangle $ transition and the detuning of the second pulse with
the $\left| 1\right\rangle \leftrightarrow \left| a\right\rangle $
transition are the same $\omega _{0a}-\omega _I=\omega _{1a}-\omega
_{II}=\Delta ,$ which creates a two-photon Raman resonance between the
levels $\left| 0\right\rangle $ and $\left| 1\right\rangle $ (a standard
technique in quantum optics [14,15]) [Fig. 1(a)], the effective interaction
Hamiltonian in the interaction picture with respect to $H_s=%
\sum_{i=0,1,a}E_i\left| i\right\rangle \left\langle i\right| $ can be
written as 
\begin{equation}
H_i=\hbar (\Omega _{0a}^Ie^{-i\Delta t+i\phi _I}\left| 0\right\rangle
\left\langle a\right| +\Omega _{1a}^{II}e^{-i\Delta t+i\phi _{II}}\left|
1\right\rangle \left\langle a\right| )+h.c..
\end{equation}
In above, $\omega _{ij}=(E_j-E_i)/\hbar $ is the transition frequency
between the levels $\left| i\right\rangle $ and $\left| j\right\rangle ,$
and $\Omega _{ij}^k=\left| \frac{\lambda _k}{2\hbar }\left\langle i\right|
\Phi \left| j\right\rangle \right| \widetilde{\Phi }_{\mu w}^k$ is the Rabi
flopping frequencies between the levels $\left| i\right\rangle $ and $\left|
j\right\rangle $ (on resonance) generated by the $k$th pulse (here, $%
i,j=0,1,a;$ $i\neq j;$ $\widetilde{\Phi }_{\mu w}^k=$ $\int_S{\bf B}%
_0^k\cdot d{\bf S;}$ and $k=I,II$).

To simplify our presentation, we now replace $\Omega _{0a}^I$ and $\Omega
_{1a}^{II}$ of Eq. (3) by $\Omega _I$ and $\Omega _{II},$ respectively. From
Eq. (3), it is easy to show that when the condition $\Delta >>\Omega _{I,II}$
is satisfied (i.e., large detuning), the Hamiltonian (3) reduces to the
following effective Hamiltonian after eliminating the intermediate level $%
\left| a\right\rangle $ 
\begin{eqnarray}
H_{eff} &=&-\hbar [\frac{\Omega _I^2}\Delta \sigma _{00}+\frac{\Omega _{II}^2%
}\Delta \sigma _{11}  \nonumber \\
&&+ge^{i\left( \phi _I-\phi _{II}\right) }\sigma ^{-}+ge^{-i\left( \phi
_I-\phi _{II}\right) }\sigma ^{+}]
\end{eqnarray}
in a rotating frame $U=e^{-i\eta }$ with $\eta =-\Delta \left( \sigma
_{00}+\sigma _{11}\right) t$, where the operators $\sigma _{00}=\left|
0\right\rangle \left\langle 0\right| ,$ $\sigma _{11}=\left| 1\right\rangle
\left\langle 1\right| ,$ $\sigma ^{-}=\left| 0\right\rangle \left\langle
1\right| $, and $\sigma ^{+}=\left| 1\right\rangle \left\langle 0\right| .$
In Eq. (4), the first two terms are ac-Stark shifts of the levels $\left|
0\right\rangle $ and $\left| 1\right\rangle ,$ which are induced by the two
microwave pulses, respectively; while the last two terms indicate that the
levels $\left| 0\right\rangle $ and $\left| 1\right\rangle $ are coupled to
each other with an effective coupling constant $g=\Omega _I$ $\Omega
_{II}/\Delta $, due to the two-photon Raman transition. It is clear that Eq.
(4) can be rewritten as 
\begin{equation}
H_{eff}=\hbar \omega _0\sigma _z-\hbar ge^{i\left( \phi _I-\phi _{II}\right)
}\sigma ^{-}-\hbar ge^{-i\left( \phi _I-\phi _{II}\right) }\sigma ^{+},
\end{equation}
where $\omega _0=(\Omega _I^2-\Omega _{II}^2)/(2\Delta )$ and $\sigma
_z=\sigma _{11}-\sigma _{00}.$

In the following, the two logical states of a SQUID qubit are represented by
the two lowest levels $\left| 0\right\rangle $ and $\left| 1\right\rangle $
of the SQUID. The dynamics of the qubit is governed by the Hamiltonian (4)
or (5). Here, it should be mentioned that a term $-\hbar \Delta \left(
\sigma _{00}+\sigma _{11}\right) $ in Eq. (4) and another term $-\hbar \frac{%
\Omega _I^2+\Omega _{II}^2}{2\Delta }(\sigma _{00}+\sigma _{11})$ in Eq. (5)
have been safely omitted, because these two terms are proportional to the
identity I $=\left| 0\right\rangle \left\langle 0\right| +\left|
1\right\rangle \left\langle 1\right| $ in the qubit Hilbert space, i.e. they
bring a common phase factor to the states $\left| 0\right\rangle $ and $%
\left| 1\right\rangle $ during the time evolution.

By means of Eq. (5), one can easily obtain the time evolution of the logical
states $\left| 0\right\rangle $ and $\left| 1\right\rangle $ as follows 
\begin{eqnarray}
\left| 0\right\rangle &\rightarrow &(\cos \sqrt{g^2+\omega _0^2}\,t+i\frac{%
\omega _0}{\sqrt{g^2+\omega _0^2}}\sin \sqrt{g^2+\omega _0^2}\,t)\left|
0\right\rangle  \nonumber \\
&&\ +i\frac{ge^{-i\left( \phi _I-\phi _{II}\right) }}{\sqrt{g^2+\omega _0^2}}%
\sin \sqrt{g^2+\omega _0^2}\,t\left| 1\right\rangle ,  \nonumber \\
\left| 1\right\rangle &\rightarrow &i\frac{ge^{i\left( \phi _I-\phi
_{II}\right) }}{\sqrt{g^2+\omega _0^2}}\sin \sqrt{g^2+\omega _0^2}\,t\left|
0\right\rangle  \nonumber \\
&&\ +(\cos \sqrt{g^2+\omega _0^2}\,t-i\frac{\omega _0}{\sqrt{g^2+\omega _0^2}%
}\sin \sqrt{g^2+\omega _0^2}\,t)\left| 1\right\rangle .
\end{eqnarray}
In particular, for the case of $\omega _0=0,$ i.e., $\Omega _I=\Omega _{II}$%
, or $\omega _0\ll g,$ i.e., $\delta \Omega \ll \Omega _I\Omega _{II}/%
\overline{\Omega }$ (here, $\delta \Omega =\Omega _I-\Omega _{II}$ and $%
\overline{\Omega }=\frac{\Omega _I+\Omega _{II}}2$), the state rotation is
given by 
\begin{eqnarray}
\left| 0\right\rangle &\rightarrow &\cos gt\left| 0\right\rangle
+ie^{-i\left( \phi _I-\phi _{II}\right) }\sin gt\left| 1\right\rangle , 
\nonumber \\
\left| 1\right\rangle &\rightarrow &ie^{i\left( \phi _I-\phi _{II}\right)
}\sin gt\left| 0\right\rangle +\cos gt\left| 1\right\rangle .
\end{eqnarray}
Eq. (6) or (7) shows that the intermediate level $\left| a\right\rangle $ is
not populated during the time evolution. This is due to the fact that the
level $\left| a\right\rangle $ is not involved in the above Hamiltonian (4)
or (5).

Based on Eq. (7), one can see that (i) A flipping between $\left|
0\right\rangle $ and $\left| 1\right\rangle $ (i.e., $\pi $-rotation) can be
implemented by setting $t=\pi /(2g)$; (ii) The Hadamard transformation $%
\left| 0\right\rangle \rightarrow \left( \left| 1\right\rangle +\left|
0\right\rangle \right) /\sqrt{2}$ and $\left| 1\right\rangle \rightarrow
\left( \left| 1\right\rangle -\left| 0\right\rangle \right) /\sqrt{2}$ can
be achieved with $t=\left( \pi /2\right) /(2g)$ and $\phi _I-\phi _{II}=\pi
/2$; (iii) More importantly, the single-qubit phase shift gate $\left|
0\right\rangle \rightarrow e^{-i\left( \phi _I-\phi _{II}\right) }\left|
0\right\rangle $ and $\left| 1\right\rangle \rightarrow e^{i\left( \phi
_I-\phi _{II}\right) }\left| 1\right\rangle $ can be realized by a two-step
operation: setting the operation time $t=\pi /(2g)$ with an arbitrary $\phi
_I-\phi _{II}$ for the first step and $t=\pi /(2g)$ and $\phi _I-\phi
_{II}=\pi $ for the second step, respectively. Thus, a complete set of
single-qubit arbitrary operations can be achieved by combining the
qubit-state rotation (7) with a single-qubit phase shift [13].

It is interesting to note that the gate operations described above can be
much faster than the conventional gate operations via resonant Rabi
oscillations between the qubit levels $\left| 0\right\rangle $ and $\left|
1\right\rangle $ [Fig. 1(a)]. For the sake of concreteness, let's choose a
SQUID with the following parameters: $\beta _L=1.20,$ $Z_0=50\Omega ,\omega
_{LC}=5\times 10^{11}$rad/s (i.e., $C=40$ fF$,$ $L=100$ pH$,$ $I_c=3.95$ $%
\mu $A), and $\Phi _x=-0.501\Phi _0,$ where $\beta _L\equiv 2\pi LI_c/\Phi
_0 $, $Z_0\equiv \sqrt{L/C}$, and $\omega _{LC}=1/\sqrt{LC}$ are,
respectively, the SQUID's potential shape parameter, characteristic
impedance, and characteristic frequency. For simplicity, denote $\phi
_{ij}\equiv \left\langle i\right| \Phi \left| j\right\rangle /\Phi _0.$ A
simple numerical calculation shows $\phi _{0a}=8.4\times 10^{-5},$ $\phi
_{1a}=5.4\times 10^{-5},$ and $\phi _{01}=7.9\times 10^{-7}$. Based on the
values of these coupling matrix elements, assuming the amplitudes of the
microwave pluses for the above three-level gate and the conventional
two-level gate on the same order, we have $\Omega _I\simeq \Omega
_{II}\simeq 10^2\Omega _{01}$. Thus, if we set the detuning $\Delta =10$ max 
$\left( \Omega _I,\Omega _{II}\right) ,$ the effective Rabi-flopping
frequency $g$ in equation (7) and the Rabi-flopping frequency $\Omega _{01}$
in the conventional gate operation have the relationship $g\simeq 10\Omega
_{01}$, i.e., the gate speed in the present scheme can be as about ten times
as that in the conventional scheme. Here, we should point out that setting $%
\Delta =10\Omega _I$ or $10\Omega _{II}$ is reasonable, because if $\Delta
=10\Omega _I,$ the maximum population of the level $\left| a\right\rangle $
would be about $\Omega _I^2/(\Omega _I^2+100\Omega _I^2)\simeq 0.01$ or $1\%$
during the gate operations.

Some points may need to be addressed here. Firstly, since both the $\Omega
_I $ and $\Omega _{II}$ couplings are far-off resonance, no transition
between any two levels is induced by a single pulse. Secondly, in order to
avoid two-photon Raman transitions between the levels $\left| 0\right\rangle 
$ and $\left| a\right\rangle $ or the levels $\left| 1\right\rangle $ and $%
\left| a\right\rangle $, the conditions $\omega _I\pm \omega _{II}$ $\neq
\omega _{0a},\omega _{1a}$ must be satisfied. Fortunately, the condition $%
\omega _I-\omega _{II}$ $\neq \omega _{0a},$ $\omega _{1a}$ is ensured
because of $\omega _{0a},\omega _{1a}\gg \omega _{01}=$ $\omega _I-\omega
_{II}$. On the other hand, the condition $\omega _I+\omega _{II}$ $\neq
\omega _{0a},\omega _{1a}$ can be easily satisfied by adjusting the SQUID's
level structure [see Fig. 1(a)]. In fact, all of the conditions required can
be readily achieved experimentally by adjusting microwave frequencies, or
level spacings of the SQUID qubit.

Recently, geometric quantum computing has been paid much attention because
apart from its fundamental interest, geometric phase is intrinsically
fault-tolerant to certain types of computational errors due to its geometric
property [3,16-20]. On one hand, several basic ideas of geometric computing
based on Berry phase have been proposed by using NMR [16], superconducting
electron box [17], or trapped ions [18]; and an experimental realization of
the conditional adiabatic phase shift has been reported with NMR technique
[16]. On the other hand, geometric computing based on Aharonov and Anandan
(A-A) phase [21] is currently getting considerable attention and several
proposals have been also presented recently [3,19,20]. In A-A phase
geometric manipulation, if the external field is perpendicular to the
evolution path, there is no dynamic phase accumulated in the whole process;
thus no extra operation is required to eliminate the dynamic phase. In
contrast, geometric manipulation based on Berry phase always involves the
dynamic phase and an extra operation is needed to remove its adverse effect
[3]. In the following, we show explicitly how to perform an arbitrary
rotation and a phase shift for a SQUID qubit via the A-A phase geometric
manipulation [22].

We start by rewriting the Hamiltonian (5) as 
\begin{eqnarray}
H_{eff} &=&\hbar \omega _0\sigma _z-\hbar g\cos \left( \phi _I-\phi
_{II}\right) \sigma _x-\hbar g\sin \left( \phi _I-\phi _{II}\right) \sigma _y
\nonumber \\
\ &=&\stackrel{\rightharpoonup }{B}\cdot \stackrel{\rightharpoonup }{\sigma }%
,
\end{eqnarray}
where the fictitious field $\stackrel{\rightharpoonup }{B}=\left\{ -\hbar
g\cos \left( \phi _I-\phi _{II}\right) ,-\hbar g\sin \left( \phi _I-\phi
_{II}\right) ,\hbar \omega _0\right\} $ and $\stackrel{\rightharpoonup }{%
\sigma }$ are the Pauli operators. To see how the logical states $\left|
0\right\rangle ${\bf \ }and $\left| 1\right\rangle $ are rotated via
geometric means, let us first show how the eigenstates $\left| \pm
\right\rangle $ of $\sigma _y$, defined by $\sigma _y\left| \pm
\right\rangle =\pm \left| \pm \right\rangle ,$ will undergo a cyclic
evolution and obtain a A-A phase under the following operations: (i) turn on
the two microwave pulses $I$ and $II$, which have $\phi _I-\phi
_{II}=(2n+1)\pi $ ($n$ is an integer)$.$ After a time $t=\pi /(2\sqrt{\omega
_0^2+g^2}),$ the state $\left| +\right\rangle $ rotates around the
fictitious field $\stackrel{\rightharpoonup }{B}=\left\{ \hbar g,0,\hbar
\omega _0\right\} ,$ from $\left| +\right\rangle $ in the \^e$_y$ direction
to $\left| -\right\rangle $ in the -$\,$\^e$_y$ direction along curve ABC on
the Bloch sphere [see Fig. 1(b)]. (ii) Change the two-pulse phase difference
to $\phi _I-\phi _{II}=2n\pi .$ After another time $t=\pi /(2\sqrt{\omega
_0^2+g^2}),$ the state $\left| -\right\rangle $ rotates back to $\left|
+\right\rangle $ around $\stackrel{\rightharpoonup }{B}=\left\{ -\hbar
g,0,\hbar \omega _0\right\} $ along curve CDA on the Bloch sphere. In the
above operations, the trace of the state vector encloses an area on the
Bloch sphere. After the above cyclic evolution, the state $\left|
+\right\rangle $ becomes $e^{-i2\theta }\left| +\right\rangle $ with a A-A
geometric phase $-2\theta ,$ where $2\theta =2\arctan (\omega _0/g)$ is the
solid angle subtended by the area of ABCDA. It should be mentioned that
because the state vector is always perpendicular to the effective magnetic
field during the above operations, no dynamical phase is accumulated during
the evolution. In a similar way, the state $\left| -\right\rangle $ will
become $e^{i2\theta }\left| -\right\rangle $ with an accumulated A-A phase
of $2\theta $ after the above operations. Finally, based on $\left|
0\right\rangle =-i\left( \left| +\right\rangle -\left| -\right\rangle
\right) /\sqrt{2}$ and $\left| 1\right\rangle =\left( \left| +\right\rangle
+\left| -\right\rangle \right) /\sqrt{2},$ one can see that at the end of
the above operation, the logical states $\left| 0\right\rangle $ and $\left|
1\right\rangle $ are rotated as 
\begin{eqnarray}
\left| 0\right\rangle &\rightarrow &\cos \alpha \left| 0\right\rangle -\sin
\alpha \left| 1\right\rangle ,  \nonumber \\
\left| 1\right\rangle &\rightarrow &\sin \alpha \left| 0\right\rangle +\cos
\alpha \left| 1\right\rangle ,
\end{eqnarray}
where $\alpha =2\theta $ is within the range of $\left[ 0,2\pi \right] ,$
i.e., (9) accomplishes a rotation with angle $\alpha $.

For the case of $\Omega _I=$ $\Omega _{II},$ i.e., $\omega _0=0,$ the
Hamiltonian (8) reduces to $H_{eff}=\stackrel{\rightharpoonup }{B}\cdot 
\stackrel{\rightharpoonup }{\sigma }$ with $\stackrel{\rightharpoonup }{B}%
=\left\{ -\hbar g\cos \left( \phi _I-\phi _{II}\right) ,-\hbar g\sin \left(
\phi _I-\phi _{II}\right) ,0\right\} .$ To implement a single-qubit phase
shift gate, one can first apply two pulses with a certain phase difference $%
\delta \phi _0.$ After a time $t=\Delta \pi /\left( 2\Omega _I\Omega
_{II}\right) ,$ the logical states $\left| 0\right\rangle $ and $\left|
1\right\rangle $ rotate around the effective magnetic field $\left\{ -\hbar
g\cos \delta \phi _0,-\hbar g\sin \delta \phi _0,0\right\} $ to $\left|
1\right\rangle $ and $\left| 0\right\rangle ,$ respectively. In the second
step of the operation, one changes the phase difference to $-\delta \phi _0.$
After another time $t=\Delta \pi /\left( 2\Omega _I\Omega _{II}\right) $,
the logical states rotate around the effective magnetic field $\left\{
-\hbar g\cos \delta \phi _0,\hbar g\sin \delta \phi _0,0\right\} $ and
return to the original states $\left| 0\right\rangle $ and $\left|
1\right\rangle .$ After this cyclic evolution, the logical states acquire
geometric A-A phases as $\left| 0\right\rangle \rightarrow e^{i\beta }\left|
0\right\rangle $ and $\left| 1\right\rangle \rightarrow e^{-i\beta }\left|
1\right\rangle ,$ where $\beta =\pi -2\delta \phi _0$ is the solid angle
subtended by the area of the loop. According to [13], combining this phase
shift gate with the qubit state rotation (9) constitutes a complete set of
single-qubit arbitrary operations.

One significant point needs to be made here. For the rotation gate, it is
not necessary to change {\it suddenly} the two-pulse phase difference after
the first-step operation. Imagine that after the first step, the two pulses
are turned off for a time interval $\delta t$ in order to adjust the phase
difference to $-\delta \phi _0$. In this case, by a simple calculation, we
find that after the whole operation, the two logical states evolve into 
\begin{eqnarray}
\left| 0\right\rangle &\rightarrow &\cos \alpha \left| 0\right\rangle
-e^{-i\left( E_1-E_0\right) \delta t/\hbar }\sin \alpha \left|
1\right\rangle ,  \nonumber \\
\left| 1\right\rangle &\rightarrow &\sin \alpha \left| 0\right\rangle
+e^{-i\left( E_1-E_0\right) \delta t/\hbar }\cos \alpha \left|
1\right\rangle ,
\end{eqnarray}
where the relative phase $e^{-i\left( E_1-E_0\right) \delta t/\hbar }$
between $\left| 0\right\rangle $ and $\left| 1\right\rangle $ is induced due
to the free evolution of the logical states during the period of adjusting
the phase difference. However, from Eq. (10), one can see that if the time
interval satisfies $\delta t=2n\pi /(\frac{E_1-E_0}\hbar )$ ($n$ is an
integer), the qubit-state rotation described by Eq. (10) is the same as that
given by Eq. (9) exactly.

In summary, we have explicitly shown how to perform an arbitrary-operation
gate with a SQUID qubit. The scheme has three distinct advantages:\ (i) The
intermediate level is unpopulated during the gate operation, thus
decoherence due to energy relaxation from the level $\left| a\right\rangle $
is minimized; (ii) No tunneling between the qubit levels $\left|
0\right\rangle $ and $\left| 1\right\rangle $ is required, therefore the
decay from the level $\left| 1\right\rangle $ can be made negligibly small
via increasing the potential barrier between the qubit levels; (iii) For the
3-level gates described here, stronger microwave field can be used to
achieve faster operation than the conventional 2-level gates. More
interestingly, as shown above, an arbitrary quantum logic via global A-A
phase geometric means can be achieved in a SQUID qubit, by extending the
present method. In addition, the method can also be applied to any other
type of qubits with a $\Lambda $-type level configuration. The present
proposal provides new approaches to demonstrate a single-qubit arbitrary
gate with superconducting devices, and we hope that the proposed approaches
will stimulate further theoretical and experimental activities in this area.

Before we conclude, it should be noticed that recently, M. H. S. Amin {\it %
et al.} [23] have proposed a similar scheme for obtaining a complete set of
one-qubit gate with a SQUID qubit. In their proposal, an arbitrary gate
operation is achieved based on a kind of rotation that is performed over a
half period of Rabi oscillation by applying microwave to induce transition
to the intermediate level (a higher state). As addressed in Ref. [23], the
probability of finding the system in the intermediate level is zero again
after a half of Rabi period. However, under their assumption of small
detuning, population in the intermediate level is finite during the
evolution, resulting in higher probability of spontaneous decay. Finally, we
should point out that qubit operations via adiabatic passage [24] in general
are rather slow and require precise control over the amplitude of external
fields.

This work was supported in part by the National Science Foundation
(EIA-0082499), and AFOSR (F49620-01-1-0439), funded under the Department of
Defense University Research Initiative on Nanotechnology Program (DURINT)
and by the ARDA.

\begin{center}
{\large Figure Captions\\}
\end{center}

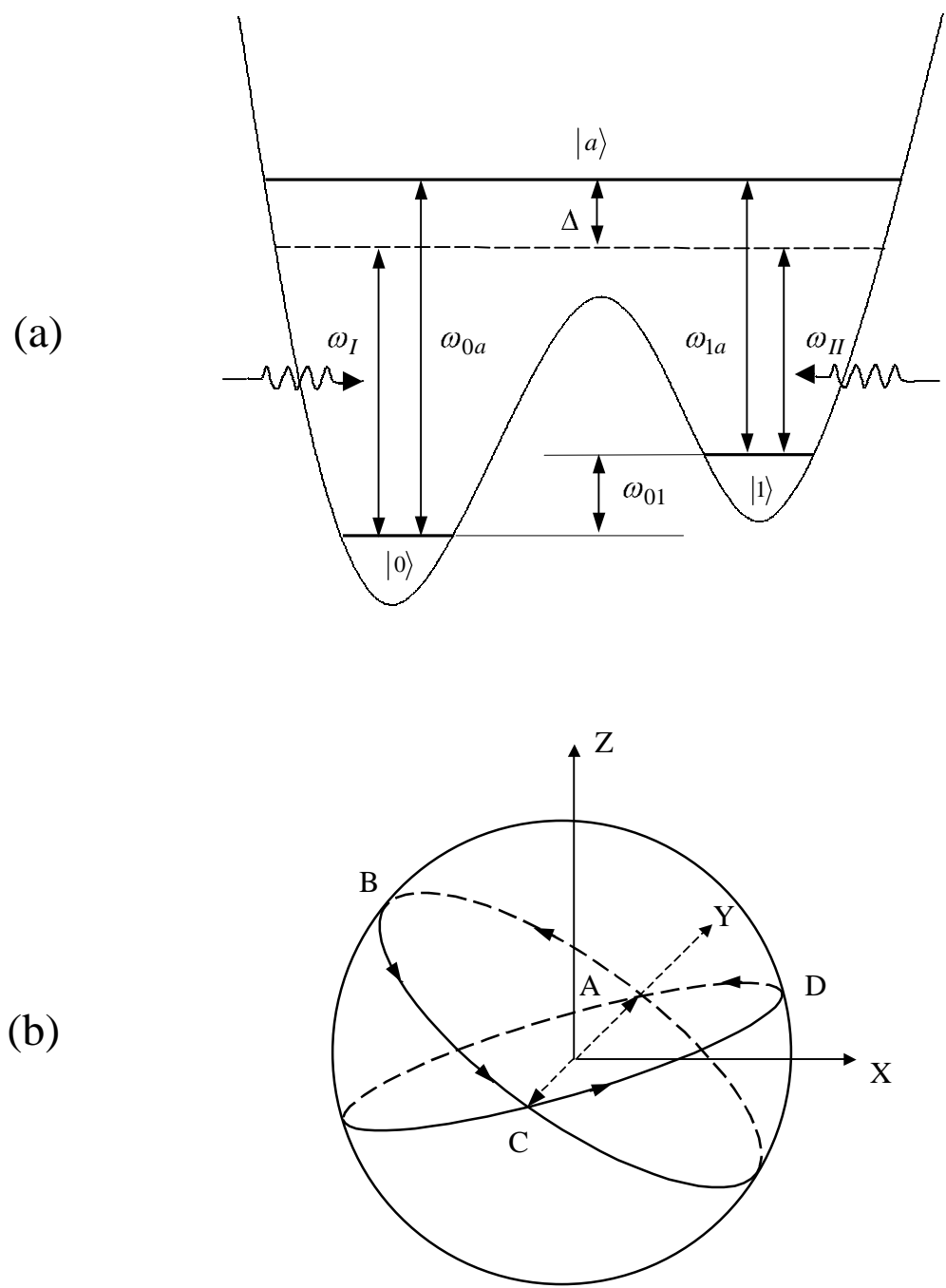
FIG. 1. (a) Simplified level diagram of an rf SQUID with $\Lambda $-type
three levels $\left| 0\right\rangle ,\left| 1\right\rangle $ and $\left|
a\right\rangle $. (b) The path evolution of the SQUID qubit initially in the
state $\left| +\right\rangle .$


\begin{references}
\bibitem{s22}  We should mention that the geometric manipulation in our
scheme is much in the same principle as Ref. [3], which is for a different
type of qubit. In our case the two logic states of a qubit are represented
by the two lowest energy fluxoid states of the SQUID, while the two logic
states of the qubit in [3] are the two charge states of a Cooper pair box
differing by one Cooper pair.
\end{references}
\end{document}